\documentclass[aip,rsi,amsmath,amssymb,
 reprint,]{revtex4-1}
\usepackage{graphicx}
\usepackage[breaklinks]{hyperref}
\usepackage{mathtools}
\usepackage{color}

\begin{document}
\title{Multi-pulse fitting of Transition Edge Sensor signals from a near-infrared continuous-wave source}

\author{Jianwei~Lee}
\affiliation{Center for Quantum Technologies,
             National University of Singapore,
             3 Science Drive 2, Singapore, 117543}

\author{Lijiong~Shen}
\affiliation{Center for Quantum Technologies,
             National University of Singapore,
             3 Science Drive 2, Singapore, 117543}
\affiliation{Department of Physics,
             National University of Singapore,
             2 Science Drive 3, Singapore, 117542}

\author{Alessandro~Cer{\`e}}
\affiliation{Center for Quantum Technologies,
             National University of Singapore,
             3 Science Drive 2, Singapore, 117543}

\author{Thomas~Gerrits}
\affiliation{National Institute of Standards and Technology (NIST),
             325 Broadway, Boulder,Colorado 80305, USA}

\author{Adriana~E.~Lita}
\affiliation{National Institute of Standards and Technology (NIST),
             325 Broadway, Boulder,Colorado 80305, USA}

\author{Sae~Woo~Nam}
\affiliation{National Institute of Standards and Technology (NIST),
             325 Broadway, Boulder,Colorado 80305, USA}

\author{Christian~Kurtsiefer}
\email[]{christian\_kurtsiefer@nus.edu.sg}
\affiliation{Center for Quantum Technologies,
             National University of Singapore,
             3 Science Drive 2, Singapore, 117543}
\affiliation{Department of Physics,
             National University of Singapore,
             2 Science Drive 3,  Singapore, 117542}
%\thanks{}
\date{\today}
\begin{abstract}
    Transition-edge sensors (TES) are photon-number resolving
    calorimetric spectrometers with near unit efficiency.
    Their recovery time, which is on the order of microseconds, limits the number
    resolving ability and timing accuracy in high photon-flux conditions.
    This is usually addressed by pulsing the light source or discarding overlapping
    signals, thereby limiting its applicability.
    We present an approach to
    assign detection times to overlapping detection events
    in the regime of low signal-to-noise ratio, as in the case of TES detection of near-infrared radiation.
    We use
    a two-level discriminator,
    inherently robust against noise, to coarsely locate pulses in time,
    and timestamp individual photoevents by fitting to a heuristic model.
    As an example,
    we measure the second-order time correlation
    of a coherent source in a single spatial mode using a single TES detector.
    \\
    \\
    Contribution of NIST, an agency of the U.S. government, not subject to copyright.
\end{abstract}

% \pacs{37.10.Gh,% Atom traps and guides
% 42.50.Ct,       % Quantum description of interaction of light and matter;
%                 % related experiments
% 32.90.+a        % Other topics in atomic properties and interactions of atoms;
%                 % with photons (restricted to new topics in section 32)
% }

\maketitle
\section{Introduction}
Transition-edge sensors are wideband photon-number resolving light detectors
that can be optimized for high quantum efficiency ($>98\%$) and to work in different regions of the electromagnetic spectrum, from soft X-rays to telecom wavelengths~\cite{Lita:2010, Fukuda:11}.
Their high single photon detection efficiency in the optical band was instrumental in one of the
recent loophole-free experimental violations of Bell's inequality~\cite{Giustina:2015}.
Absorption of a single photon by the TES generates an electric pulse response
with a fast (tens of nanoseconds) rising edge, and a relaxation with a time
constant of a few microseconds~\cite{LamasLinares:2013fk}.
Photodetection events with time separation
shorter than the pulse duration overlap and cannot be reliably identified
by threshold crossing.
To avoid this problem, TES are often used with pulsed light sources with a
repetition rate lower than few tens of kilohertz~\cite{Levine:2012by}. This may
exclude the use of TES with superb detection efficiencies from some applications.
Therefore, in this work we investigate the time discrimination for
overlapping signal pulses using a continuous-wave~(CW) light source.

Similar problems are common in high-energy physics~\cite{Marrone2006,BELLI2008512,Vencelj2009,Tambave2012, Fowler:2015ef}.
Fowler et al.~\cite{Fowler:2015ef} improved time discrimination
by considering the time derivative of the signal
to locate the steep rising edge of individual photodetection events.
In cases with high signal-to-noise ratio, such as in the
detection of high-energy photons ~$\gamma$ and X-rays (SNR $\approx$260, estimated from Ref.~\onlinecite{Fowler:2015ef}),
this approach is effective also when signals overlap.
%%%%%%%SNR estimation%%%
% Fig 4 caption: n=1 at 0.8eV
% pg 3: resolution 0.33 eV
% SNR = 0.8/0.33 = 2.4
%%%%%%%%%%%%%%%%%%%%%%%%
However,
for near-infrared (NIR) photodetection with a TES, it is necessary to filter
high frequency noise components to improve the signal-to-noise ratio
(SNR $\approx$2.4, estimated from Ref.~\onlinecite{LamasLinares:2013fk}) at the
expense of a reduced timing accuracy.

We approach the problem by separating it into two distinct phases:
an initial event identification, followed by a more accurate timing discrimination.
We identify photodetection events using a two-level discriminator.
Its resilience to noise allows us to
coarsely locate both isolated and overlapping pulses
with a moderate use of filtering,
thus retaining some of the high frequency components of the signal, useful to improve the time accuracy of subsequent operations.
For monochromatic sources, every detection event has the same energy.
We can then
estimate the number of photons for every detection region from the total pulse area, identifying the cases of overlapping events.
From the number of photons, we calculate a heuristic model function and fit it
to the signal
to recover the detection-times.

\section{Electronics and photon detection pulse}\label{sec:electronics}
\begin{figure}
\includegraphics[width=1\linewidth]{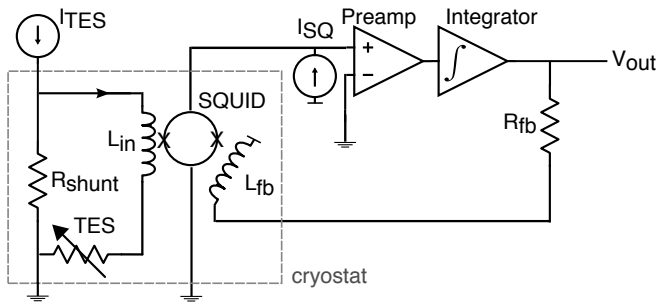}
  \caption{\label{fig:setup}
   Schematic of the TES biasing and readout electronics.
   The TES is voltage-biased by a constant current source~$I_{\text{TES}}$ through shunt resistor~$R_{\text{shunt}} \ll R_{\text{TES}}$.
   The SQUID array amplifier picks up changes in TES resistance from~$L_{\text{in}}$.
   The signal is further amplified outside of the cryostat.
   Signal feedback via~$R_{\text{fb}}$ and coil~$L_{\text{fb}}$ linearizes the SQUID response.
    }
\end{figure}

Our tungsten-based TES~\cite{Lita:08} is kept at a
temperature of 75\,mK using an adiabatic demagnetization refrigerator cryostat, and is voltage biased within its superconducting-to-normal transition in a negative electro-thermal feedback~\cite{Irwin:1995ie}.
The detection signal is inductively picked up and amplified by
a SQUID series array,
followed by further signal conditioning at room temperature
with an overall amplification bandwidth
of~$\approx$6\,MHz.
A schematic of the TES biasing and readout electronics is shown in figure~\ref{fig:setup}.
We operate the SQUID in a flux-locked loop~\cite{Drung:2007iu} to minimize low frequency
components of the noise.
To characterize the TES response, we use a laser diode centered at 810\,nm as a light source, operated in CW mode.
We control the average photon flux with a variable attenuator, then launch
the light into a fiber (type SMF28e~\cite{disclaimer}) that directs it to the
sensitive surface of the TES.

We record~10\,$\mu$s long traces with a sampling rate of $5\times10^8$~s$^{-1}$ and a 12\,bit voltage resolution.
For light at 810\,nm, the signal generated by discrete absorption processes for
each photon after the amplifier chain exhibits a rise time for a single photon
pulse of about~100\,ns, and an overall pulse duration of about~$2\,\mu$s.

We collected a total of~$4\times10^5$ traces with the TES continuously
illuminated by an attenuated laser diode.
Despite the flux-locked loop, we observe a residual voltage offset variation from trace to trace.
Therefore, for every recorded pulse trace $v_{\text{rec}}(t)$, we remove the residual baseline,
\begin{equation}
    v(t) = v_{\text{rec}}(t) - V_{M}\,,
\end{equation}
where~$V_{M}$ is the most frequently occurring value of the discretized signal~$v_{\text{rec}}(t)$ over the sampling interval.

\section{Pulse Identification}\label{sec:disc}
\begin{figure}
\includegraphics[width=1\linewidth]{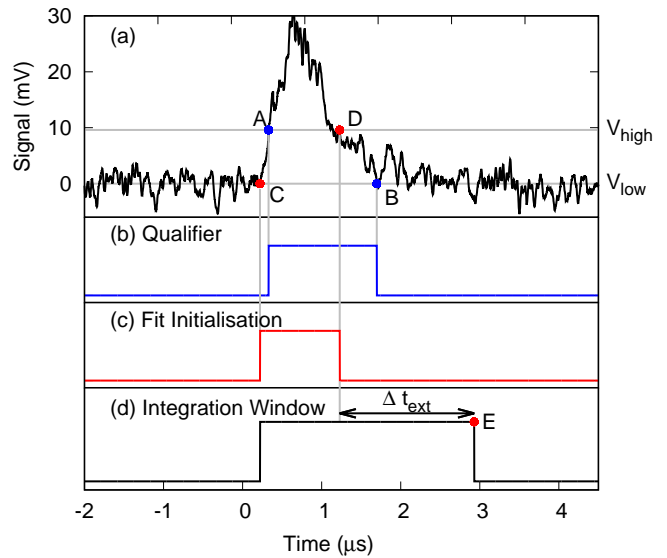}
  \caption{\label{fig:disc}
    (a) Typical TES   response
    with overlapping pulses.
    The horizontal lines show the high and low threshold settings of the
    Schmitt trigger mechanism.
    (b) Qualifying interval AB identified by the Schmitt trigger.
    (c) The interval~CD includes the rising edges of the overlapping pulses, and is used to initialize a least-square fit.
    (d) The wider interval~CE that includes the rising edge and decaying tail is used to estimate the number of photons associated with the event.
    We empirically found a reasonable energy resolution with
    Point~E obtained by extending interval~CD by $\Delta t_{ext} = 1700$\,ns.
    }
\end{figure}
In a first step, we identify the presence of an absorption process from one or
more photons in a trace, and distinguish it from background noise. This is done by a
traditional Schmitt trigger mechanism~\cite{Schmitt:1938}, implemented via
discriminators at two levels:
a qualifier flag is raised when the signal passes threshold~$V_{\text{high}}$ (figure~\ref{fig:disc}(a),~point A)
and lowered by the first subsequent crossing of threshold~$V_{\text{low}}$
(point B).

\begin{figure}
\includegraphics[width=1\linewidth]{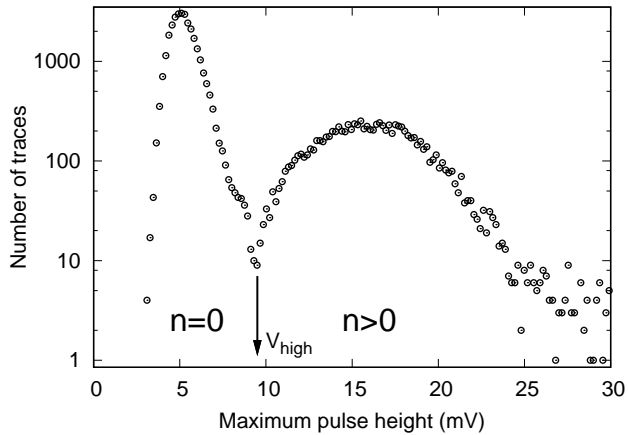}
    \caption{\label{fig:height_histo}
        Histogram of maximum pulse heights for~$4\times10^5$ traces.
        The two distributions correspond to traces with ($n>0$) and without ($n=0$) photodetection events.
        We use the minimum between the two distributions to set the threshold $V_{\text{high}}$ of the discriminator.}
\end{figure}

In order to
minimize the number of false events,
we estimate~$V_{\text{high}}$
using
a histogram of maximum pulse heights for~$4\times10^4$ traces,
shown in figure~\ref{fig:height_histo}.
The distribution has two distinct peaks, with one around 5\,mV corresponding to traces without any detection event~($n=0$),
and another one starting from 9.5\,mV onwards corresponding to traces with at least one detection event~($n>0$).
We choose $V_{\text{high}}$ to the minimum between the two peaks (9.5\,mV), and
$V_{\text{low}}$ to 0\,mV.

An expected timing accuracy for single photon events that can be extracted
from the TES response
would be given by the RMS~noise (about 1.75\,mV), and the
steepest slope of the response (0.11(9)\,mV/ns, estimated from an average of
10\%-90\% transitions of an ensemble of pulses) to be about
16\,ns. However, a simple threshold detection of the leading edge does not
work if pulses start to overlap.

More precise timing information of a photodetection event is obtained from a
least square fit to the signal using a displaced standard pulse. To efficiently
initialize this fit, we do not directly use the qualifier window AB for two
reasons:
first, it contains only a fraction of the leading edge belonging to the
earlier pulse that contains most of the timing information, and second, it includes a large portion of the decaying tail unassociated with the onset of photodetection.
The time window CD derived from the same discriminator levels ensures the
inclusion of the first leading edge, and is also shorter.

Similarly, we derive an integration time window from the qualifier
window to determine the pulse integral, from which we extract the photon
number of a quasi-monochromatic light source.
As a starting point, we choose point~C for the integration to capture the
rising slope of a pulse, and extend the time~D by a fixed amount~$\Delta
t_{ext}$ to point~E to capture the tail of the response signal
(figure~\ref{fig:disc}(d)).
We found that it is more reliable to extend point~D by a fixed time to capture
the tail of the signal rather than to reference the end of the integration window to
point~B. This is because the signal-to-noise ratio around~B is low, leading to a large
variation of integration times.
We empirically find that~$\Delta t_{ext}=1700$\,ns gives a good signal-to-noise ratio of the pulse integral.

\section{Photon Number Discrimination}\label{sec:number_discrimination}
To determine the number of photons in each trace, we assume that the detection
and subsequent amplification have a linear response,
so that the integral of each signal is proportional to the absorbed energy~\cite{Cabrera2000509}, resulting in a discrete distribution of the
areas of the signals. This distribution is spread out by noise, and we have to
use an algorithm to extract the photon number in presence of this noise.

\begin{figure}
\includegraphics[width=1\linewidth]{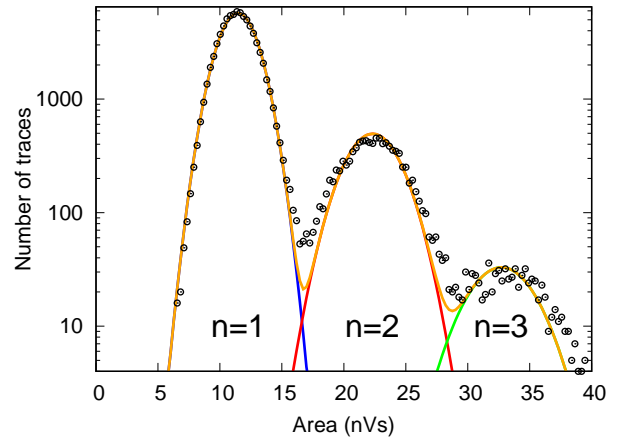}
    \caption{\label{fig:area_histo}
        Distribution of pulse areas~$H(a)$.
        For every trace that triggers the two-levels discriminator, the area is calculated within the region~CE.
      The continuous lines are Gaussian fits for the~$n=1$ (blue),~$n=2$ (red), and~$n=3$ (green) area distributions,
      and their sum (orange).
    }
\end{figure}

For this, we first compute the pulse area~$a=\int_{t_C}^{t_E}|v(t)|$ for every
qualified trace within region CE.
Figure~\ref{fig:area_histo} shows a histogram of pulse areas from the qualified traces out of all the~$4\times10^5$ acquired.
The distribution shows three resolved peaks
that suggest having been caused by $n=1,2,3$~photons being absorbed by the TES.

One can fit the histogram in figure~\ref{fig:area_histo}
to a sum of three normalized Gaussian peaks~$g_n(a; a_n, \sigma_n)$\,,
\begin{equation}\label{eq:peakfit}
    H(a) = \sum_{n=1}^{3}h_n\,g_n(a| a_n, \sigma_n)\,,
\end{equation}
where each Gaussian peak is characterized by an average area~$a_n$ and
width~$\sigma_n$. The ratio $a_2/a_1=1.95$ indicates that the TES response to
photon energies of 1 and 2 photons is approximately linear.

We identify thresholds $a_{n-1,n}$ as the values that minimize the overlap
between distributions~$g_{n-1}(a| a_{n-1}, \sigma_{n-1})$ and~$g_n(a| a_n,
\sigma_{n})$. With this, we assign a number of detected photons~$n$ by
comparing the area of every trace to thresholds~$a_{n-1,n}$ and~$a_{n,n+1}$.

The continuous nature of the light source with a fixed power level makes it
difficult to assign a number of photons per qualified signal, as the
integration window varies from pulse to pulse, and detection events
may occur at random times in the respective integration
windows. Heuristically, however, one could even replace the individual event
numbers $h_n$ in Eq.~\ref{eq:peakfit} by a Poisson distribution,
\begin{equation}
h_n=N p(n|\bar{n})\,,
\end{equation}
where $\bar{n}$ is an average photon number, $p(n|\bar{n})$ the Poisson
coefficient, and $N$ is the total number of traces. For the data shown in
figure~\ref{fig:area_histo}, this would lead to an average photon number of
$\bar{n}\approx0.3$ per integration time interval.

\section{Determining the detection-times of overlapping pulses}\label{sec:fitting}
The difficulty of assigning a photon number to light detected from a CW source
can be resolved if one treats the first detection process of light following
the paradigm of wideband photodetectors in quantum
optics~\cite{Glauber:1963}. 
As TES are sensitive over a relatively wide optical
bandwidth, the corresponding time scale of the absorption process 
is much shorter than
the few microseconds of the TES thermal recovery~\cite{Gerrits:2016kb}.
% explicit recovery time reference: 
% Springer International Publishing Switzerland 2016
% R.H. Hadfield and G. Johansson (eds.), Superconducting Devices
% in Quantum Optics, Quantum Science and Technology,
% DOI 10.1007/978-3-319-24091-6_2 
Then, the
signal would correspond to a superposition of responses to individual
absorption processes, 
which may happen at times closer than the
characteristic pulse time.

To recover absorption times of individual absorption events in a trace
of~$N$ overlapping pulses, where~$N$ is determined with the pulse area method
outlined in the previous section, we fit the TES response signal~$v(t)$
to a heuristic model~$v_N(t)$ of a linear combination of single-photon
responses~$v_1(t)$,
\begin{equation} \label{eqn:model}
    v_N(t | \{t_i, A_i\})) = \sum_{i=1}^N\,A_i\,v_1(t-t_i)\,,
\end{equation}
where~$A_i$ is the amplitude
and~$t_i$ the detection time of the $i$-th pulse. While the TES response
to multi-photon events is not strictly linear, this model will give a reasonably
good estimation of the timing for single photon absorption events.

\subsection{Single photon pulse model}\label{sec:1ph_model}
    \begin{figure}
    \centering
    \includegraphics[width=1\linewidth]{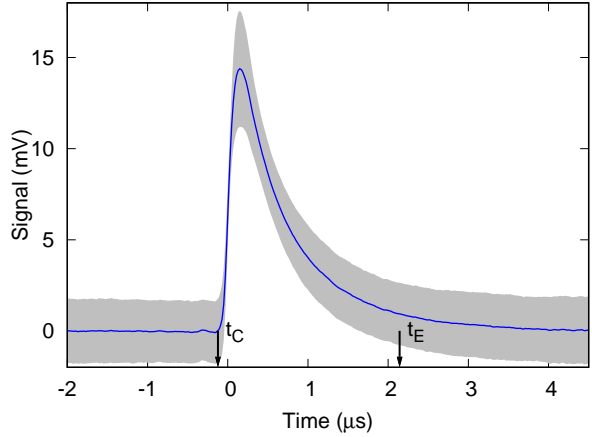}
      \caption{
      \label{fig:intrinsic_jitter}
        Solid line: average response of the TES and amplification to a single
        absorption.
        We use a Schmitt trigger to identify the region between
        $t_C$ and $t_E$.
        Grey region: one standard deviation in the observed ensemble of~$n=1$ traces.
       }
    \end{figure}

We obtain a model for the single photon response $v_1(t)$ of the TES and its
signal amplification chain for the fit in Eq.~\ref{eqn:model} by selecting
$N_1=10^4$ single photon traces from the measurement shown in
figure~\ref{fig:area_histo}, and averaging over them. The averaging process
eliminates the noise from individual traces, and retains the detector response.

Signal photon events can happen at any time within the sampling window.
It is necessary to
align
these detection events to average the traces.
We assign a detection time to the $i$-th trace $v_1^{(i)}(t)$ by recording the time~$t_i$ corresponding to the maximum of~$dv_1^{(i)}(t)/dt$.
We use a Savitzky-Golay filter (SGF)
to reduce the high frequency components~\cite{Savgol:1964};
the SGF replaces every point with the result of a linear fit to the subset of adjacent~41 points.

We also reject clear outlier traces by limiting the search for~$t_i$
to the time interval~CD.
The remaining~$N_1$ traces are then averaged by synchronizing them according
to their respective~$t_i$ and to obtain the single-photon response~$v_1(t)$,
\begin{equation}\label{eq:ph1-model}
  v_1(t)=\frac{1}{N_1}\sum_{i=1}^{N_1}v_1^{(i)}(t + t_i)\,.
\end{equation}
The result is shown in figure~\ref{fig:intrinsic_jitter}, together with a noise
interval derived from the standard deviation of~$N_1$ single photon
traces.
The model demonstrates an average rise time of 116\,ns
from 10\% to 90\% of its maximum height.
The relaxation time (1/$e$) of 635\,ns corresponds to detector thermalization~\cite{Lita:2010cx}.

\subsection{Time-tagging via least-square fitting}
\begin{figure}
    \includegraphics[width=1\linewidth]{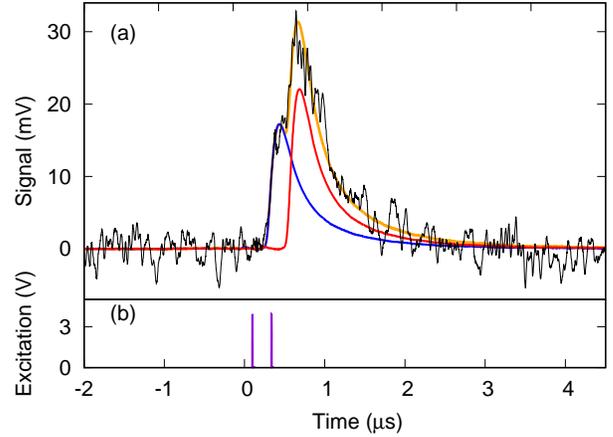}
    \caption{\label{fig:fit}
    (a) Fit of a two-photon signal with the heuristic function described in the main text.
    Black line: measured TES response after removing the vertical offset.
    Orange line: fit to Eq.~(\ref{eqn:model}), with two single photon
    components separated in time (blue and red line).
    (b) Electrical pulse pair separated by 239\,ns sent to the LD illuminating the TES.}
\end{figure}

For every qualified trace, we assign a number of photons~$N$ according to the calculated area,
and
fit it using Eq.~(\ref{eqn:model}).
The fit has~$2\,N$ free parameters:
detection times~$t_i$ and amplitudes~$A_i$, with~$i = 1\dots N$.
We bound~$t_i$ to the range CD (figure~\ref{fig:disc}(c)), and restrict
the sum of~$A_i$ to be consistent with the thresholds obtained from the area
distribution
\begin{equation}
    \frac{a_{N-1,N}}{\int_{t_C}^{t_E} |v_1(\tau)| \,d\tau}\leq
    \sum_{i=1}^N A_i \leq
    \frac{a_{N,N+1}}{\int_{t_C}^{t_E} |v_1(\tau)| \,d\tau}\,.
\end{equation}
The accuracy of the fit depends on the choice of minimization algorithm.
We used Powell's derivative-free method~\cite{Powell1964} because
the presence of noise tends to corrupt gradient estimation~\cite{Plagianakos2002}.

\begin{figure}
%    \centerline{\includegraphics[width=1\linewidth]{figures/fit_accuracy/fit_accuracy_powell.eps}}
    \centerline{\includegraphics[width=1\linewidth]{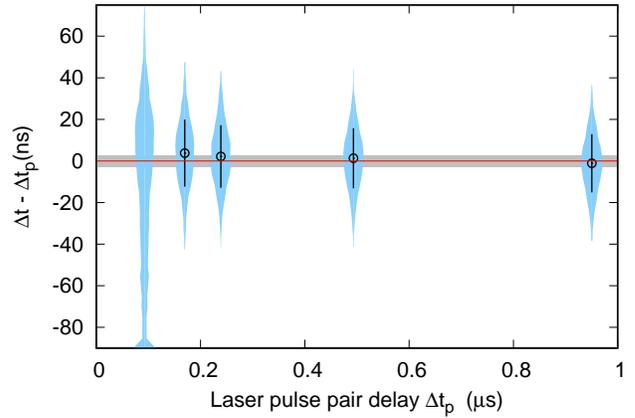}}
      \caption{\label{fig:g2fixedtime}
        Difference between the detection-time separation estimated with the fitting technique~($\Delta t$)
        and
        the delay of laser pulse pairs~($\Delta t_p$)
        for five different delays: 92\,ns, 170\,ns, 239\,ns, 493\,ns, and 950\,ns.
        Blue regions: distribution of~$\Delta t - \Delta t_p$.
        Grey region: expected range of separation for~90\% of
        single photon detections for~4\,ns long laser pulse pairs.
        Black circles: mean of the distributions with error bars corresponding to one standard deviation.
        }
\end{figure}

To verify the accuracy of the fitting algorithm for~$N=2$, we expose the TES
to pairs of short (4\,ns) laser pulses with a controlled delay~$\Delta
t_p$.
The 100\,kHz repetition rate is low enough to isolate
the TES response between consecutive laser pulse pairs.
Selecting only the traces with two photons, we have two possible cases:
(i) a two-photon event generated within one of the 4\,ns pulses
or (ii) one photon in each pulse.
We compared the TES response for five different delays~$\Delta t_p$: 92\,ns, 170\,ns, 239\,ns, 493\,ns, and 950\,ns.
Figure~\ref{fig:fit} shows an example of a measured trace where the fitting algorithm was able to distinguish between separate photodetection events at~$\Delta t_p=239$\,ns even though
it appears to be a single event because of the detector noise.
For each delay we collected~$\approx3.5 \times 10^5$ traces, and
for each trace we estimate the photodetection times using the least-square method.
In figure~\ref{fig:g2fixedtime} we summarize the distribution of time differences $\Delta t = |t_2 - t_1| $
for each delay.

Except for the shortest pulse separation,
the time differences have Gaussian distributions with standard deviations
of about 16\,ns. This matches the time accuracy expected from
the simple noise/slope estimation for the
leading edge of the single photon pulse, despite the overlapping pulses.
The average separation between the center of the distribution and the expected result, $\Delta t-\Delta t_p$, is~2(2)\,ns.
For $\Delta t_p=92$\,ns, the distribution is clearly skewed toward~$0$\,ns.
This multi-modal distribution indicates that the fit procedure is unable to distinguish two single-photon
events generated by the two separated diode pulses from two-photon events
generated within the same diode pulse.

\section{Detection-Time Separation from coherent source}
As another benchmark for the fitting technique presented in here, we extract
the normalized second order correlation function~$g^{(2)}(\Delta t)$  for
detection  events recorded with a single TES from a coherent light field.
For a light field in a coherent state, this correlation function should be
exactly 1 for all time differences~$\Delta t$~\cite{Glauber:1963}.

For this, the TES is exposed to light from a continuously running laser diode,
with an average photon number of about 0.3 per integration interval of around
3\,$\mu$s. Again, we select only two-photon traces using the methods described in section~\ref{sec:number_discrimination}, and fit the traces to the model described by equation~\ref{eqn:model} with~$N=2$.
\begin{figure}
    \includegraphics[width=1\linewidth]{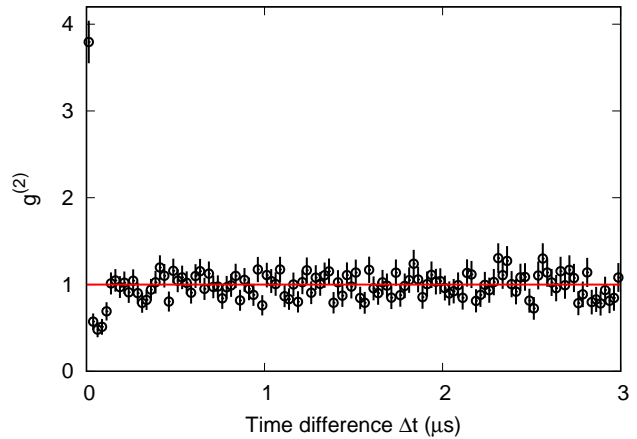}
      \caption{\label{fig:g2cw}
        Normalized second order correlation function $g^{(2)}(\Delta t)$ for   events recorded with a single TES from a coherent light field.
        Error bars indicate one standard deviation assuming Poissonian statistics, the bin size is~25\,ns.
        Solid line: expected correlation for a coherent field.
        }
    \end{figure}

Each fitted trace leads to two time values~$t_1$ and~$t_2$, which we sort into
a frequency distribution~$G^{(2)}(\Delta t)$ of time differences~$\Delta
t=t_2-t_1$.
We normalize this distribution with the distribution expected for a Poissonian
source, taking into account the finite time of our acquisition windows.
We remove single-photon traces mis-identified as two-photon traces by
filtering out traces that have a minimum estimated amplitude smaller than one half of a single photon pulse.

The resulting normalized distribution $g^{(2)}(\Delta t)$ is shown in
figure~\ref{fig:g2cw}.
For $\Delta t>150$\,ns, the correlation function is
compatible with the expected value of 1.
For shorter time differences,
the fit algorithm occasionally locks on the same detection times $t_1$ and
$t_2$, redistributing pair events to $\Delta t=0$,
resulting
in a calculated correlation then deviates from the expected behavior, including the unphysical value $g^{(2)}(\Delta t=0)>2$.
This instability region ($\Delta t<150$\,ns)
is comparable with the rise time of the average
single-photon pulse, and is consistent with the precision
indicated in figure~\ref{fig:g2fixedtime}.

\section{Conclusion}
We demonstrated a signal processing method based on a Schmitt-trigger based data
acquisition and a linear algorithm that can reliably extract both a photon
number and photodetection times from the signal provided by an optical
Transition Edge Sensor (TES) with an accuracy that is mostly limited by the
detector time jitter.

Using this method,  we successfully resolved between~$n=1, 2$ and~$3$ photons
from a CW NIR source, using the signal integral evaluated in the time interval
identified by the discriminator.
The time interval includes a greater fraction of the photodetection signal
than that considered by a single-threshold discriminator.
By considering an optimal fraction of the pulse profile,
we obtained pulse integral distributions that sufficiently resolve between photon numbers.
We note that the maximum pulse height is unsuitable for photon number discrimination of a CW~source since the maximum height depends on the photodetection times when pulses are overlapped. This is evident in figure~\ref{fig:height_histo}. 
In contrast, figure~\ref{fig:area_histo} shows that~$n=1, 2$ and~$3$ photon events are well resolved using the pulse integral, which does not depend on photodetection times.

This technique provides an alternative to photon counting
using edge detection on the differentiated signal~\cite{Fowler:2015ef}
when signal-to-noise ratio is low.

The discriminated region is then used to initialize a least-squares fit of a
signal containing two overlapping pulses to a two-photon model,
returning the amplitudes and detection-times of the individual photons.

With the available TES, we can distinguish two photodetection events within
about 200\,ns using this method.
The highest detection rate that can be processed is thus estimated to be about~$5\times10^6$~s$^{-1}$,
compared to about~$4\times10^5$~s$^{-1}$ if we were to discard overlapping
pulses.

Potential applications include the measurement of time-resolved correlation functions using the TES without the need for the spatial multiplexing of several single-photon non-photon-number resolving detectors, provided that the coherence time of the light source is larger than the timing resolution of this technique.
The order of the correlation function measured is limited only by the maximum number of photons resolvable by the TES.

\section*{Acknowledgments} This research is supported by the Singapore Ministry of Education Academic Research Fund Tier 3 (Grant No.~MOE2012-T3-1-009); by the National Research Fund and the Ministry of Education, Singapore, under the Research Centres of Excellence programme.
% This work includes contributions of the National Institute of Standards and Technology, which are not subject to U.S.~copyright.

%\bibliography{pulse_fitting.bib}

\begin{thebibliography}{22}
\bibitem [{dis()}]{disclaimer}%
  \BibitemOpen
  \href@noop {} {\bibinfo  {journal} {Certain commercial equipment, instruments
  or materials are identified in this report to foster understanding. Such
  identification does not imply recommendation or endorsement by the National
  Institute of Standards and Technology, nor does it imply that the materials
  or equipment are necessarily the best available for the purpose.}\
  }\BibitemShut {NoStop}%
\end{thebibliography}

\end{document}